\documentclass{llncs}

\usepackage{verbatim}
\usepackage{algorithm}
\usepackage{algorithmic}
\usepackage{enumerate}
\usepackage{paralist}
\usepackage{fmtcount} 
\usepackage{amsmath}

\usepackage{amssymb}
\usepackage{turnstile}

\def\True{{\it true}}
\def\False{{\it false}}
\def\cA{$\cal A$}

\begin{document}

\title{A Family of Translations for Pseudo-Boolean Constraints to CNF}
\titlerunning{Translating PB-Constraints to SAT}

\author{Amir Aavani\inst{1} \and David Mitchell \inst{1} \and Eugenia Ternovska\inst{1}
}
\authorrunning{Amir Aavani, David Mitchell, Eugenia Ternovska}


%
\institute{Simon Fraser University, Computing Science Department\\
\email{\{aaa78,mitchell,ter\}@sfu.ca} 
}

\newtheorem{myobservation}{Observation}
\newtheorem{mytheorem}{Theorem}
\newtheorem{mylemma}{Lemma}
\newtheorem{myremark}{Remark}
\newtheorem{mydefinition}{Definition}
\newtheorem{myproposition}{Proposition}
\newtheorem{myproof}{Proof}
\newtheorem{myexample}{Example}
\newcommand{\Here}[1]{\noindent\hspace{-2mm}\fbox{\fbox{\parbox{\textwidth}{#1}}}}
\newcommand{\here}[1]{{\bf *** #1 ***}}
\newcommand{\MOD}[1]{{\textrm{mod\ #1}}}
\newcommand{\VAR}[1]{{$var$\ (#1)}}

\maketitle

\begin{abstract}
  A Pseudo-Boolean constraint, PB-constraint, is a linear constraint over Boolean variables. This kind of constraints has been widely used in expressing NP-complete problems. 

This paper introduces a family of algorithms for translating Pseudo-Boolean constraints into CNF clauses. These algorithms are centered around the idea of rewriting a PB-constraint as the conjunction of a set of easier to translate constraints, we call them PBMod-constraints. The CNF produced by the proposed encoding has small size, and we also characterize the constraints for which one can expect the SAT solvers to perform well on the produced CNF. We show that there are many constraints for which the proposed encoding has a good performance.

We compared the running time of SAT solvers on the output of the proposed translation and the existing approaches.

\end{abstract}
%
\section{Introduction}\label{Introduction}

%
%
  A Pseudo-Boolean constraint (PB-constraint), which is also known as 0-1 integer linear constraint by the integer linear programming community, is a generalization of a clause. A PB-constraint is an inequality (equality) on a linear combination of Boolean literals:
\begin{center}
$\sum_{i=1}^n a_il_i\{<,\le,=,\ge,>\} b$,
\end{center}
 where $a_1,\cdots a_n$ and $b$ are constant integers and $l_1,\cdots,l_n$ are literals. The left-hand side of a PB-constraint under assignment \cA\ is equal to the sum of the coefficients whose corresponding literals are mapped to \True\ by \cA. 

  One way to build a solver which is capable of handling PB-constraints is to modify a SAT solver to support PB-constraints natively. PBS \cite{aloul2002pbs} and PUEBLO \cite{sheini2006pueblo} are examples of such solvers. The main challenge in this approach is to modify/extend all the heuristic functions used in the original SAT solver. Another approach is to replace a given PB-constraint with a logically equivalent set of clauses and then use a SAT solver to find a solution. The main benefit of the latter approach is that every SAT solver, even those which are going to be developed in future, can be plugged in to the system. Also, there are certain NP problems which can be translated into a combination of a relatively small CNF formula plus one or two PB-constraints. One can name Vehicle Routing Problem and its variations \cite{VRP}, Hamiltonian Cycle problem and Knapsack as examples of such problems. Having a {\em good} translation for PB-constraints enables both naive and expert users to use SAT solvers for attacking these problems. Most professional users encode these problems using Integer Linear Programming (ILP) tools. Unfortunately, there is no natural way to express certain sentences in an integer linear program, e.g. ``Either John or Maria is wearing a green shirt and a black hat''.



  We define a PBMod-constraint to be:
\begin{center}
$\sum_{i=1}^n a_il_i \equiv b\ (\MOD{M}),$
\end{center}

 where $a_1,\cdots a_n$ and $b$ are non-negative integers less than $M$, and $l_1,\cdots,l_n$ are literals. 
 
 In Section 3, we show that instead of translating a given PB-constraint, we can translate a set of appropriately selected PBMod-constraints. So to translate PB-constraints to SAT, we need to determine how to \textit{choose} the set of PBMod-constraints and how to \textit{translate} a PBMod-constraint to SAT. As we show in this paper, there are many PB-constraints whose unsatisfiability  can be proven by showing the unsatisfiabiliy of a PBMod-constraint. Some of our translations for PBMod-constraints allow unit-propagation to infer inconsistency if the current assignment cannot be extended to a satisfying assignment for that PBMod-constraint and hence unit-propagation can infer inconsistency for the original PB-constraint. In Section 6, it has been shown that the number of PB-constrains for which unit-propagation can infer inconsistency, given the output of proposed translations, is much larger than the other existing encodings. Also, we prove that it is impossible to translate all PB-constraints in the form $\sum a_il_i=b$ into polynomial size arc-consistent CNF unless P=CoNP.

  The structure of this paper is as follows: The next section is devoted to preliminaries and definitions. The proposed encoding is presented in Section 3 and 4. In Section 5, four existing translations (encodings) for converting a PB-constraint to CNF are described.  In Section 6, we study the performance of unit propagation on the resulting CNF of different encodings. Specifically, we describe a necessary condition on the instances for which our translation is arc-consistent, and also show that there is no polynomial size arc-consistent encoding for PB-constraint in the form $\sum a_il_i=b$ unless P=Co-NP.

\section{Background}\label{Background}
  In this section, we fix our notations and use them through the rest of this paper. Also, we define when an encoding produces a \textit{valid translation}.

\subsection{Notations}

  Let $X$ be a set of Boolean variables. A literal, $l$, is either a Boolean variable or negation of a Boolean variable and \VAR{l} denotes the variable corresponding to $l$. A clause on $X$, $C=\{l_1,\cdots,l_{m}\}$, is a set of literals such that \VAR{$l_i$}$\in X$. An assignment \cA\ to $X$ is a function that maps some variables in $X$ to either \True\ or \False. By $x\in {\cal A}^+$($x\in {\cal A}^-$) we mean that \True\ (\False) is assigned to $x$ under assignment \cA. Also, we use \cA$\ [S]$, $S\subseteq X$, as a shorthand for the assignment obtained by restricting the domain of \cA\ to the variables in $S$.

PB-constraint $Q$ on $X$ is specified as:
\begin{center}
$a_1l_1+\cdots+a_n l_n\qquad \{<,\le,=,\ge,>\}\qquad b,$
\end{center}

where each $a_i$ is the integer \textit{coefficient} of $l_i$; $b$ is an integer, called \textit{bound}, and $l_i$ is a literal, s.t., \VAR{$l_i$}$\in X$.

Assignment \cA\ to $X$ is a {\em{total assignment}} if it assigns a value to each variable in $X$, i.e., \cA$^+\cup $\cA$^-=X$. Assignment \cA\ satisfies literal $l$, ${\cal A}\models l$, if $l=x$ and $x\in$\cA$^+$\ or $l=\lnot x$ and $x\in$\cA$^-$. Assignment \cA\ satisfies clause $C=\{l_1,\cdots,l_m\}$ if there exists at least one literal $l_i$ such that \cA$\models l_i$. A total assignment falsifies clause $C$ if it does not satisfy any of its literals. An assignment satisfies a set of clauses if it satisfies all the clauses in that set. Total assignment \cA\ to $X$ satisfies a PB-constraint $Q$ on $X$, $A\models Q$ , if the value of left-hand side of $Q$ under \cA, i.e., $\sum_{i: {\cal A} \models l_i} a_i$ and that of right-hand side of $Q$ satisfies the comparison operator.

We say assignment \cA\ extends assignment $\cal B$, \cA$\supseteq\cal B$, iff both ${\cal B}^+\subseteq$\cA$^+$ and ${\cal B}^-\subseteq$\cA$^-$ hold. 

\subsection{Valid Translation}

  Here, we formalize the meaning of translation of a constraint into CNF and we use this definition to prove the correctness.

  Note that a constraint can be seen as a Boolean function which returns \True\ on assignments that satisfy the constraint and \False\ otherwise.

\begin{mydefinition}\label{ValidTranslation} Given Boolean function $F(X)$, where $X=\{x_1,\cdots,x_n\}$ is a set variables, we call the pair $\langle v,C\rangle$, where $v$ is a Boolean variable, $C=\{C_1,\cdots,C_m\}$ is a set of clauses on $X\cup Y \cup \{v\}$ and $Y$ is a set of (auxiliary) propositional variables, a {\em{valid translation}} if $C$ is satisfiable and for every total assignment \cA\ to $X\cup Y\cup\{v\}$ that satisfies $C$, \cA\ satisfies $F(X)$ iff it maps $v$ to \True, i.e., $C \not\models \bot, $ and:

\begin{center}
$C,{\cal A}[X]\models v\Leftrightarrow F(X|_{\cal A})=\True.$
\end{center}

\end{mydefinition}

Intuitively, $C$ describes the relation among input variables, $x\in X$, auxiliary variables, $y\in Y$, and $v$. The truth value of $v$ is the same as the truth value of $F(X)$ under all assignments which satisfy $C$. 

\begin{myobservation} Let $Y=F (X)$ be an $n$-input $m$-output Boolean function. Function $F$ can be described using $m$ Boolean functions ($f_1(X),\cdots,f_m(X)$) where $f_i$ computes the $i$-th output of $F$. Then a valid translation for $F$ can be constructed using valid translations for $f_i$'s. Let $\langle v_i,C_i\rangle$ be valid translation for $f_i$, for $i=1\cdots m$. Pair $\langle V,C\rangle$ is a valid translation for $F$, where $V=\{v_1,\cdots,v_m\}$ and $C=\cup C_i$.
\end{myobservation}

  In \cite{bailleux2009new}, a translation is defined to be just a set of clauses. It is easy to verify that these two definitions are equivalent. 

  It worths mentioning that our definition of a valid translation is not limited to PB-constraints.

\begin{myexample} Let $Q$ be the following PB-constraint which is not satisfiable. Based on definition \ref {ValidTranslation}, the pair $\langle v,\{C_1\}\rangle$ where $C_1=\{\lnot v\}$ is a \em{valid translation} for $Q$.
$$Q:2x_1+4\lnot x_2=3.$$
\end{myexample}

\subsection{Tseitin Transformation}
  The usual method for transforming a propositional formula to CNF is by the method of Tseitin\cite{tseitin1968complexity}.  In this transformation, a fresh propositional variable is created to represent the truth value of each subformula of the given formula. Let $\psi_1,\psi_2,\psi$ be three such subformulas and $x,y,z$ be the associated propositional variables to $\psi_1,\psi_2$ and $\psi$, respectively. The transformation works as follows:

\begin{compactenum}
  \item $\psi=\psi_1 \lor \psi_2:$ produce the following three clauses $\{\lnot z,x,y\},\{z,\lnot x\},\{z,\lnot y\}$: (i.e., $z\Leftrightarrow x\lor y$),
  \item $\psi=\psi_1 \land \psi_2:$ produce the following three clauses $\{\lnot z,x\},\{\lnot z,x\},\{z,\lnot x,\lnot y\}$: (i.e., $z\Leftrightarrow x\land y$),
  \item $\psi=\lnot \psi_1:$ produce the following two clauses $\{\lnot z,\lnot x\},\{z,x\}$ (i.e., $z\Leftrightarrow \lnot x$),
  \item $\psi=v $, where $v$ is a propositional variable: produce the following two clauses $\{\lnot z,v\},\{z,\lnot v\}$ (i.e., $z\Leftrightarrow v$).
\end{compactenum}

\subsection{Canonical Form}
  Let consider the following PB-constraint: 

\begin{equation}\label{GeneralForm}
  a_1l_1+\cdots+a_n l_n = b,
\end{equation}

\noindent where all constant integers ($a_1\cdots a_n$ and $b$) are positive integers. We show that every PB-constraint can be rewritten as a PB-constraint in form of \ref{GeneralForm}.

\begin{mydefinition}\label{EqConstraint} 
  Constraints $Q_1$ on $X$ and $Q_2$ on $Y\supseteq X$ are \textit{equivalent} iff for every satisfying assignment \cA\ for $Q_1$, there exists at least one expansion of \cA\ to $Y$ satisfying $Q_2$, and for every total assignments \cA\ to $X$ which does not satisfy $Q_1$, all possible expansions of \cA\ to $Y$ falsifies $Q_2$. 
\end{mydefinition}

\begin{myobservation} \label{PBLIsGeneral}Let $n\ge 1$. The following PB-constraints are equivalent. 
\begin{compactenum}
\item\label{PBGE} $\sum_{i=1}^{n} a_il_i>=b,$
\item\label{PBG} $\sum_{i=1}^{n} a_il_i>b- 1,$
\item\label{PBLE} $\sum_{i=1}^{n} -a_il_i<=-b,$
\item\label{PBL} $\sum_{i=1}^{n} -a_il_i<1-b.$
\end{compactenum}
\end{myobservation}

\begin{myobservation} \label{PosPBLIsGeneral}Let $m$ and $n$ be such that $1\le m\le n$. Then (\ref{PBLPosNeg}) and (\ref{PBLPos}) are equivalent.
\begin{compactenum}
\item\label{PBLPosNeg} $\sum_{i=1}^{m-1} a_il_i+ a_ml_m+\sum_{i=m+1}^{n}a_il_i<b,$
\item\label{PBLPos}  $\sum_{i=1}^{m-1} a_il_i- a_m\lnot l_m+\sum_{i=m+1}^{n}a_il_i<b-a_m.$
\end{compactenum}
\end{myobservation}

Observation \ref{PBLIsGeneral} and observation \ref{PosPBLIsGeneral} imply that every PB-constraint whose comparison operator is in $\{\le,<,>,\ge\}$ can be rewritten as an equivalent PB-constraint with positive coefficients in the following form:

\begin{equation}\label{PreGeneralForm}
 \sum_{i=1}^{n} a_il_i< b.
\end{equation}

If the right-hand side of (\ref{PreGeneralForm}), $b$, is less than or equal to zero, no assignment satisfies the constraint, i.e., the pair $\langle v,\{\{\lnot v\}\}\rangle$ can be used a valid translation for it. It is not hard to observe that if we have a PB-constraint whose left-hand side is $1$, pair $\langle v, \{\{\lnot v,\lnot l_1\},\cdots,\{\lnot v, \lnot l_n\},\{v, l_1,\cdots,l_n\}\}\rangle$\footnote{The clauses in $C$ corresponds to $v\Leftrightarrow \lnot l_1\land \cdots \lnot l_n$.} is a valid translation for that constraint. 

\begin{myproposition} Let $n\ge 1$, $a_i\ge0,b> 1$ and $B=\lfloor \log_2 b\rfloor$. Also, assume the variables $y_i$ are newly introduced Boolean variables. Then (\ref{PBL2}) and (\ref{PBEQ}) are equivalent.
\begin{compactenum}
\item\label{PBL2} $\sum_{i=1}^{n} a_il_i< b$
\item\label{PBEQ} $\sum_{i=1}^{n} a_il_i+ \sum_{i=0}^{B}2^iy_i=b-1$
\end{compactenum}
\end{myproposition}

In conclusion, every PB-constraints can be rewritten as an equivalent normalized PB-constraint in form \ref{GeneralForm}. So, if we know how to find a valid translation for a PB-constraint in form (\ref{GeneralForm}), we can find a valid translation for every PB-constraint, as well.

\subsection{Unit Propagation}
  {\em Unit propagation} (UP) is a mechanism used by SAT solvers to accelerate the search process. Whenever the current partial assignment maps all but one of the literals in a clause to false, the value of the remaining literal should be true if the instance is satisfiable. A similar situation can happen for PB-constraints, i.e., given a partial assignment \cA\ and a PB-constraint $Q$ on $X$, there might be a variable that takes the same value in all satisfying expansion of \cA. So, the value for that variable is forced. 

  Given an assignment \cA, the PB-constraint $Q$ on $X$ can be transformed to an equivalent PB-constraint $Q'$ on $Y$ such that all the variables in $Y$ are unassigned under \cA:
 \begin{eqnarray}
 \nonumber  Q: \sum a_il_i=b & \\
 \nonumber  Q': 0+\sum_{i:var (l_i)\in Y}&a_il_i=b-\sum_{i:{\cal A}\models l_i}a_i
 \end{eqnarray} 
 The terminology used here is an adaptation of what has been used in \cite{bailleux2009new}. A translation for the given constraint $Q$ is {\em UP-detectable} if UP infers inconsistency whenever there is no assignment that satisfies $Q$. A translation for the given constraint $Q$ is {\em UP-inferable} if, for any literal $l$, UP infers the value of $l$ whenever $l$ takes the same value in all satisfying solutions to $Q$. More formally, let $\langle v,C\rangle$ be a valid-translation for $Q$ on $X$. The pair $\langle v,C\rangle$ is UP-detectable if $Q\models \bot \Leftrightarrow C\land v\sststile{UP}{} \bot$. It is UP-inferable if $Q\models l\Leftrightarrow C\land v\sststile{UP}{} l$. A translation for $Q$ is {\em generalized arc-consistent}, or simply arc-consistent, if it is both UP-detectable and UP-inferable. An encoding is arc-consistent if it produces an arc-consistent translation for all possible input constraints.

\section{Proposed Method}\label{ProposedMethod}

In this section, we focus on describing how our proposed approach works on the PB-constraints which are in the following form:

\begin{equation}\label{NormalPB}
\sum_{i=1}^n a_ix_i=b,
\end{equation}
\noindent where all constants are positive integers and $x_i\ne x_j$, for all $i\ne j$.

  Let a normal PBMod-constraint be an equation in the following form: 

\begin{equation}\label{NormalPBInMod}
\sum_{i=1}^n a_ix_i\equiv b\ (\MOD M),
\end{equation}
\noindent where $0\le a_i<M$ for all $1\le i\le n$ and $0\le b<M$. Total Assignment \cA\ is a solution to a PBMod-constraint iff the value of left-hand side summation under \cA\ minus the value of right-hand side of the equation, $b$, is a multiple of $M$.

\begin{mydefinition}
  PBMod-constraint $Q[M]:\sum a'_ix_i\equiv b' (\MOD{M})$ is called to be the {\em conversion} of PB-constraint $Q:\sum a_ix_i=b$, modulo $M$ iff:
 \begin{enumerate} 
    \item $a'_i= a_i\ \MOD{M},$
    \item $b'= b\ \MOD{M}.$
 \end{enumerate} 
\end{mydefinition}

  One can verify that each solution to PB-constraint $Q$  is also a solution to all its conversions module $M$, $Q[M]$, $M\ge 2$. Also, for appropriately large values of $M$, each solution to $Q[M]$ is a solution to $Q$. So, for the appropriate values of $M$, the two constraints have the same set of solutions. Our goal is to select the value of $M$ such that translating  the corresponding PBMod-constraint is easier than translating the original PB-constraint.

\begin{mylemma} \label{BoundOnM} For any PB-constraint $Q:\sum a_ix_i=b$, if $M$ satisfies $M> S=\sum a_i$, PBMod-constraint $Q[M]$ and PB-constraint $Q$ have exactly the same set of solutions, i.e., any assignment either satisfies both equations or neither of them.
\end{mylemma}
{\noindent\bf Proof} It is obvious that if \cA\ is a solution for $Q$, \cA\ satisfies $Q[M]$, too. Now, let's \cA be a solution (a satisfying assignment) for $Q[M]$. The value of left-hand side of $Q[M]$ under \cA\ should be an integer in the form $b+k*M$ for some $k\ge 0$. As we have $0\le b+k*M\le \sum a_i< M$, we can infer that $k$ should be zero and so the sum of left-hand side of $Q[M]$ under \cA\ is exactly equal to $b$.

\begin{mylemma}\label{FactoringM} Let $Q:\sum a_ix_i=b$ be a PB-constraint. Also, let $M_1$ and $M_2$ be two integers and $M_3=\textrm{lcm}\ (M_1, M_2)$. Assume $S_j$ is the set of assignments satisfying $Q[M]$ when $M=M_j$, for $j=1,2$ and $3$. We have:
$$S_3=S_1 \cap S_2.$$
\end{mylemma}
 {\noindent\bf Proof}The proof of this Lemma is very similar to the proof of the following statement (which can be found in any number theory book, as an exercise): Let $M_1, M_2$ and $M_3$ be three integers s.t. $M_3=\textrm{lcm}\ (M_1, M_2)$. Then for any integer $x$ and $y$ we have:

$ x \equiv y\ (\textrm{mod}\ M_1) \land x\equiv y\ (\textrm{mod}\ M_2) \Leftrightarrow x\equiv y\ (\textrm{mod}\ M_3).$


Lemma~\ref{FactoringM} tells us that in order to find the set of answers to a PBMod-constraint modulo $M_3=\textrm{lcm}(M_1,M_2)$, one can find the set of answers to two PBMod-constraints (modulo $M_1$ and $M_2$) and return their intersection.

\begin{myproposition}\label{DescribingAnswer}
Let ${\tt M}=\{M_1,\cdots,M_m\}$ be a set of $m$ positve integers. The set of assignments satisfying $Q:\sum a_ix_i=b$ is exactly the same as the set of assignments satisfying all the $m$ PBMod-constraints, $Q[M_1], Q[M_2], \cdots, Q[M_m]$ if $\textrm{lcm}(M_1,\cdots,M_m)>S=\sum a_i$. 
\end{myproposition}

\begin{mytheorem}\label{MainTheorem} Let $Q:\sum a_ix_i=b$ be a PB-constraint. Assume we have access to a translation oracle which produces a valid translation for every PBMod-constraint. Let ${\tt M }=\{M_1,\cdots,M_m\}$ be as described in Prop. \ref{DescribingAnswer}, and the pair $\langle v_k,C_k\rangle$ be a \textit{valid translation} for $Q[M_k]$ obtained using the translation oracle. Then, pair $\langle v, C\rangle$, where $C=\cup_k C_k \cup C'$ and $C'$ is the set of clauses describing $v\Leftrightarrow (v_1\land v_2\cdots\land v_m)$, is a \textrm{valid translation} $Q$.
\end{mytheorem} 

Theorem \ref{MainTheorem} can be proved by a straightforward application of Lemma \ref{BoundOnM} and Proposition \ref{DescribingAnswer}.

  We know $\textrm{lcm}(2,\cdots,k)\ge 2^{k-1}$, \cite{farhi2009new}, so set ${\tt M^N}=\{2,\cdots,\lceil \log \sum a_i\rceil+1\}$ can be used as the set of modulos for encoding $Q:\sum a_ix_i=b$. 

  Another candidate for set ${\tt M}$ is subset of prime numbers. One can enumerate the prime numbers and add them to the set of modulos, ${\tt M^P}$, until their multiplication exceeds $S$, i.e., to select ${\tt M^P}$ to be $\{2,3,...,P_m\}$. The next proposition gives us an estimation for the size of set ${\tt M^P}$ as well as the maximum value in ${\tt M^P}$.

\begin{myproposition} \label{Bounds} Let ${\tt M^P}$ be the set of primes less than or equal to $P_m$ (assume $P_m$, itself, is a prime number) such that 

$$\prod_{p\in \tt M^P}p\ge S.$$ 
Then:
\begin{enumerate}
  \item $m=|{\tt M^P}|= \theta (\frac{\ln S}{\ln \ln S})$.
  \item $P_m< \ln S$.
\end{enumerate}
\end{myproposition}
Proof of this proposition can be found in the appendix \ref{Proposition7Proof}. 

The number of modulos, i.e., the size of $\tt M$, can be reduced if we choose larger modulos. One way to do so is to select the set of modulos to be ${\tt M^{PP}}=\{{P_i}^{n_i}: P_i \textrm{ is i-th prime number and } {P_i}^{n_i-1}\le \log S\le {P_i}^{n_i}\}$. So, we have fewer modulos while each modulos is not too big. 

\begin{myproposition} \label{BoundsOnPP} Let ${\tt M^{PP}}=\{{P_i}^{n_i}: P_i \textrm{ is i-th prime number and } {P_i}^{n_i-1}\le \ln S\le {P_i}^{n_i}\}$ be such that 
$$\prod_{M\in \tt M^{PP}}M\ge S.$$ 
Then:
\begin{enumerate}
  \item $m=|{\tt M^{PP}}|\le \frac{\ln S}{\ln \ln S}$\ , 
  \item $\max_{M\in M^{PP}}M= \ln S$\ .
\end{enumerate}
\end{myproposition}

{\noindent\bf Proof}
\begin{enumerate}
  \item $S\le \prod_{m\in \tt m^{PP}}M\le (\ln S)^m$
   $\Rightarrow \frac{\ln S}{\ln \ln S}\le m$\ .
   \item it comes from the construction of ${\tt M^{PP}}$. 
 \end{enumerate}
 

Note that ${\tt M^N}$, ${\tt M^P}$ and ${\tt M^{PP}}$ are just three possible sets of modulos. Given PB-constraint $Q$, there are many other candidates for the set of modulos.


It is worth mentioning that the size of description of PB-constraint $Q: \sum a_ix_i=b$ is $\theta(n\log a_{Max})$ where $n$ is the number of literals (coefficients) in the constraint and $a_{Max}$ is the maximum value of coefficients. The size of description of PBMod-constraint $Q[M]$ is $\theta(n\log M)$ where $n$ is the number of literals (coefficients) in the constraint. So, if we can come up with a translation for $Q[M]$ which produces a CNF with $O(n^{k_1} M^{k_2})$, for some constants $k_1$ and $k_2$, clauses/variables (which is exponential in its input size), we have translated the PB-constraints into CNF using a polynomial number of variables (clauses, literals) with respect to the size of representation of the original PB-constraint. Several such translations are described in the next section.

\section{Encoding For Modular Pseudo-Boolean Constraints}\label{ModularPB}

In this section, we describe how a PBMod-constraint in the format of Equation (\ref{PBMod-Constraint}), where $0\le a_i, b <M $, can be translated into CNF. Remember that our ultimate goal is not to translate PBMod-constraints but to translate PB-constraints. 

\begin{equation}\label{PBMod-Constraint}
  \sum_{i=1}^n a_i l_i = b~(\MOD M).
\end{equation}

\subsection{Translation Using DP}\label{TranslatingUsingDP}
The translation presented here encodes PBMod-constraints using a Dynamic Programming approach. Auxiliary variable $D_m^l$ is defined inductively as follows:

\[\hspace{10mm} D_{m}^{l}=\ \ \left\{
  \begin{array}{l l}
    \mbox{$\top$} & \quad \mbox{$l$ and $m$ are both zero};\\
    \mbox{$\bot$} & \quad \mbox{$l=0$ and $m> 0$};\\
    \mbox{$(D_{(m-a_{l})\MOD{M}}^{l-1} \land x_{l})\lor (D_{m}^{l-1}\land \lnot x_{l})$} & \quad Otherwise.
  \end{array}
\right. \]

This encoding is similar to translation through BDD, described in \cite{Een2006translating}. Using a top-down approach, starting from $D_b^n$, for describing the Tseitin variables usually generates a smaller CNF. 

In this encoding, auxiliary variable $D_m^l$ describes the necessary and sufficient condition for satisfiability of subproblem $\sum_{i=1}^l a_ix_i\equiv m\ (\MOD{M})$. 

\begin{myproposition} Let $D=\{D_{m}^l\}$ and $C$ be the clauses which are used to describe the variables in $D$. Then, the pair $\langle D_b^n,C\rangle$ is \textit{valid translation} for (\ref{PBMod-Constraint}).
\end{myproposition}

Adding the following clauses helps unit propagation to infer more facts:
\begin{compactenum}
  \item For each $l, m_1, m_2$, where $m_1\le m_2$: $\{\lnot D_{m_1}^l, \lnot D_{m_2}^l\}$. This clause asserts that $\sum_{i=1}^l a_ix_i$, modulo $M$, cannot be evaluated as both $m_1$ and $m_2$.
  \item For each $l$: $\{D_{m}^l|m=0\cdots M-1\}$. This clause asserts that $\sum_{i=1}^l a_ix_i$, modulo $M$, is among $0, 1, \cdots, M-1$.
\end{compactenum}

\begin{myproposition} We can use the following set of clauses to describe the relation among $D_{m}^l$, $D_{m-a_l}^{l-1}$, $D_m^l$, $x_l$:

\begin{enumerate}
  \item If both $D_{m-a_l}^{l-1}$ and $x_l$ are True, we should have $D_m^l$ is True, i.e., $\{\lnot D_{m-a_l}^{l-1}, \lnot x_l, D_m^l\}$;
  \item If $D_{m}^{l-1}$ is True and $x_l$ is False, we should have $D_m^l$ is True, i.e., $\{\lnot D_{m}^{l-1}, x_l, D_m^l\}$;
  \item If both $D_{m}^{l}$ and $x_l$ are True, we should have $D_{m-a_l}^{l-1}$ is True, i.e., $\{\lnot D_{m}^{l}, \lnot x_l, D_{m-a_l}^{l-1}\}$;
  \item If $D_{m}^{l}$ is True and $x_l$ is False, we should have $D_m^{l-1}$ is True, i.e., $\{\lnot D_{m}^{l}, x_l, D_m^{l-1}\}$;
  \item At most one of $D_{0}^l,\cdots,D_{M-1}^l$ can be True: $\{\lnot D_i^l, \lnot D_j^l\}$ ($0\le i< j< M$; 
  \item At least one of $D_{0}^l,\cdots,D_{M-1}^l$ is True: $\{D_0^l,D_1^l,\cdots,D_{M-1}^l\}$; 
\end{enumerate}

  Using this set of clauses results in an encoding for PBMod-constraints with the following property:

 Given partial assignment \cA, if there is no total assignment satisfying $C$ and extending \cA\ which maps $\sum_{i=1}^a a_ix_i$ to $m$, then unit propagation infers \False as the value for variable $D_{a}^{m}$. 

\end{myproposition}

\subsection{Translation Using DC}\label{TranslatingUsingDC}

  The translation presented here resembles a Divide and Conquer approach. Variable $D_a^{s,l}$ is defined inductively as follows:

\[D_{m}^{s,l}=\left\{
  \begin{array}{l l}
    \mbox{$\top$} & \quad \mbox{$m$ and $l$ are both zero};\\
    \mbox{$\bot$} & \quad \mbox{$l=0$ and $m\ne 0$};\\
    \mbox{$x_s$} & \quad \mbox{$l=1$ and $m\ne 0$ and $a_s=m$};\\
    \mbox{$\lnot x_s$} & \quad \mbox{$l=1$ and $m= 0$ and $a_s\ne 0$};\\
    \mbox{$\bot$} & \quad \mbox{$l=1$ and $m\ne 0$ and $a_s\ne m$};\\
    \mbox{$\top$} & \quad \mbox{$l=1$ and $m=a_s= 0$};\\
    \mbox{$\bigvee_{a'=0}^{M-1} (D_{(m-a'\textrm{ mod } M)}^{s,l\slash 2} \land D_{a'}^{s+l\slash 2,l\slash 2} )$ } & \quad Otherwise.
  \end{array}
\right. \]

Here, $D_{a}^{s,l}$ describes the necessary and sufficient condition for satisfiability of subproblem $\sum_{i=s}^{s+l-1} a_ix_i\equiv a\ (\MOD{M})$.

\begin{myproposition} 
Let $D=\{D_{a}^{s,l}\}$ and $C$ be the clauses which are used to describe the variables in $D$. Then, pair $\langle D_b^n,C\rangle$ is a \textit{valid translation} for (\ref{PBMod-Constraint}).
\end{myproposition}

Similar to translating using DP, by adding the following clauses, we can boost the performance of unit propagation for this translation, too.
\begin{compactenum}
  \item For each $s, l, m_1, m_2$, where $m_1\le m_2$: $\{\lnot D_{m_1}^{s,l}, \lnot D_{m_2}^{s,l}\}$.
  \item For each $s, l$: $\{D_{m}^{s,l}|m=0\cdots M-1\}$.
\end{compactenum}





\subsection{Translation Using Sorter}
  An $n$-bit Boolean sorter is an $n$-input $n$-output Boolean function $\langle y_1,\cdots,y_n\rangle=Sort(x_1,\cdots,x_n)$ satisfying the following constraints:
  
  \begin{compactenum}
    \item If $y_i=1$, then $y_j=1$ for all $j<= i$,
    \item The number of \True\ input variables is the same as the number of \True\ output variables, i.e., $|\{i:x_i=\top\}|=|\{i:y_i=\top\}|$.
  \end{compactenum}

  In unary representation, the numerical value of a bit-vector is the number of bits set to \True. Bit-vector $V_i=\langle x_i,\cdots,x_i\rangle$, where $|V_i|=a_i$ represents either $0$ or $a_i$, depending on the value of $x_i$. It is straightforward to see that $\sum a_ix_i=m$ is eqisatisfiable with the conjunction of the following three conditions:
  \begin{compactenum}
    \item $\langle y_1,\cdots,y_S\rangle=Sort (V)$, where $V=\langle x_1,\cdots,x_1,x_2,\cdots,x_2,\cdots,x_n\cdots,x_n\rangle$ is a bit vector and each $x_i$ occurs $a_i$ times in $V$ and $S=\sum a_i$,
    \item $y_m=\True$,
    \item $y_{m+1}=\False$.
  \end{compactenum}

  The above construction can be used to generate a valid translation for a given PBMod-constraint. Let $C_{Sorter}$ be the set of clauses describing the relation between input variables $V$, output variables $Y=\langle y_1,\cdots,y_n\rangle$ and auxiliary variables for a sorter. Then, pair $\langle v,C\rangle$ is a valid translation for (\ref{PBMod-Constraint}) where $C= C_{Sorter}\cup C_v$, and $C_v$ is the set of clauses describing 
  $$v\Leftrightarrow \bigvee_{j\equiv b'\ \MOD{M}}y_{j}$$
  
  A sorter network can be constructed either a sorting network, or BDD encoding.

\begin{myproposition} 
Let $C_{Sorter}$ be the set of clauses describing a sorter and $C_v$ be the set of clauses describing $v\Leftrightarrow \bigvee_{j\equiv b'\ \MOD{M}}y_{j}$. Then, the pair $\langle v,C\cup C_{Sorter}\rangle$ is a \textit{valid translation} for (\ref{PBMod-Constraint}).
\end{myproposition}
 
\subsection{Translation Using Cardinality Constraints}
  Let a cardinality constraint be as what we have described in constraint on a set of Boolean variables which restricts the number of True variables in the set. It can be seen that a cardinality constraint is a special case of PB-constraints where all coefficients are one:

\begin{equation}\label{CardinalityConstraint}
C:x_1+\cdots+x_n= b.
\end{equation}

Essentially, (\ref{CardinalityConstraint}) asserts that a satisfying assignment for $C$ should map exactly $b$ literals out of the literals in set $\{x_1,\cdots,x_n\}$ to \True. There are many approaches to produce a valid translation for a cardinality constraint, see \cite{Aavani2010Encoding}.

  Having a PBMod-constraint in form (\ref{PBMod-Constraint}), it can be rewritten as the following constraint:
\begin{equation}
\sum_{i=0}^{M-1}\#(\{x_j\ |\ a_j\ \MOD{M}= i\})*i\equiv\ b'\ (\MOD{M}),
\end{equation}
\noindent where $\#(\{y_1,\cdots,y_m\})$ represents the number of literals mapped to \True.




\begin{myproposition}\label{ArcConsistencyDPPBMod}  Unit-propagation infers inconsistency in the generated CNF of BDD translation iff the PBMod-constraint is unsatisfiable. UP infers the value of an input variable, $x_i$, iff that variable takes a unique value in all solutions of the input PBMod-constraint. If the PBMod-constraint has exactly one solution, UP is able to infer all input variables values.
\end{myproposition}
The proof of Proposition (\ref{ArcConsistencyDPPBMod}) is essentially the same as the proof for arc consistency of BDD encoding.

 \begin{mytheorem} Using BDD encoding as the translation oracle in Theorem \ref{MainTheorem}, one can translate the PB-constraint $Q:\sum a_il_i=b$ into a CNF with $n\sum P_i\le nmP_m\le n\log S\log S\le n(\log n+\log a_{Max})^2$ variables, $O(n(\log n+ \log a_{Max})^2)$ clauses and $O(n(\log n+ \log a_{Max})^2)$ literals. 
\end{mytheorem}
  Until now, we described how a PB-constraint can be translated into a series of PBMod-constraint and how a PBMod-constraint can be translated into CNF. In example \ref{Example1}, we demonstrate the procedure of converting a PB-constraint to CNF.

\begin{myexample}\label{Example1} Consider the following PB-constraint. For this case, we have $S=15$ and $P=\{2,3,5\}$.
$$ 1 x_1+ 2 x_2+ 3x_3+ 4x_4+5x_5= 7 $$
Let $\langle v_2,C_2\rangle$, $\langle v_3,C_3\rangle$ and $\langle v_5,C_5\rangle$ be valid translations for the following PBMod-constraints, respectively:
$$ 1 x_1+ 0 x_2+ 1x_3+ 0x_4+1x_5= 1 (\MOD 2)$$
$$ 1 x_1+ 2 x_2+ 0x_3+ 1x_4+2x_5= 1 (\MOD 3)$$
$$ 1 x_1+ 2 x_2+ 3x_3+ 4x_4+0x_5= 2 (\MOD 5)$$
Then, $\langle v,C_2\cup C_3 \cup C_5 \cup C'\rangle$, where $v$ is a new variable and $C'$ is the set of clauses necessary to describe $v\Leftrightarrow v_2\land v_3\land v_5$. 
\end{myexample}

 Note that every encodings for PB-constraints can directly be converted to an encoding for PBMod-constraints using the following observation:
$$\sum a_i l_i=b (\MOD M) \Leftrightarrow \exists k\ 0\le k\le Max: \sum a_il_i=b+ k*M$$
where $Max=\lfloor\frac{\sum a_i}{M}\rfloor< \lfloor \frac{n* (M-1)}{M}\rfloor<n$ as the left-hand side is an integer in range $[0\cdots \sum a_i]$. 

We know that every integer in range $[0\cdots Max]$ can be encoded using $\log (Max+1)$-bits. So, the following two constraints are equivalent, i.e., every solution to one of them can be uniquely converted to a solution to another one.
\begin{eqnarray}
\label{PBModToPB-PBMod}\sum a_i l_i=b (\MOD M)& \\
\label{PBModToPB-PB}\sum a_i l_i- 1k_0- 2^1 k_1&- \cdots- 2^{\lfloor \log (Max+1)\rfloor}k_{\lfloor \log (Max+1) \rfloor}=b
\end{eqnarray}

 So, instead of encoding PBMod-constraint~\ref{PBModToPB-PBMod}, one can encode the normalized version of PB-constraint~\ref{PBModToPB-PB} using any encoding which produces a valid translation for PB-constraints.  

  In particular, if we use the {\em Totalizer based} encoding, \cite{bailleux2009new}, in the above approach, we get an encoding encoding for PBMod-constraints whose CNF has at most $n^4\log n \log M$ clauses, $n^3\log n \log M$ auxiliary variables and $n^4\log n \log A$ literals. And then, we will have an encoding for PB-constraints which produces a CNF with $n^4\log n* (\log n+ \log a_{Max})$ clauses, $n^3\log n*(\log n+ \log a_{Max})$ auxiliary variables and $n^4\log n(\log n+ \log a_{Max})$ literals. But the resulting encoding for PBMod-constraints will not be arc consistent, because as we show in the next section, totalizer based encoding is not arc-consistent, for certain PB-constraints.

%

\section{Previous Work}\label{PreviousWork}



 The existence of a polynomial size arc-consistent encoding for PB-consistent in form $\sum a_ix_i<b$ was an open question until very recently. Bailluex et al. developed an arc-consistent polynomial size translation for these constraints \cite{bailleux2009new}. Although all kinds of PB-constraints can be written as conjunction of at most two constraints in the form $\sum a_il_i<b$, arc-consistency is not preserved for PB-constraints in the form $\sum a_il_i=b$. Moreover, in section \ref{GAC}, we prove there cannot be a polynomial size arc-consistent encoding for all possible PB-constraints in form $\sum a_il_i=b$ unless P= CoNP.

\subsection{Arc-consistent Encodings}

  \subsubsection{Translation through BDD}

  This approach is similar to the dynamic programming solution for solving the subset-sum problem. For every possible pair $i$ and $j$ where $0\le i\le n$, $0\le j\le b$, a fresh Tseitin variable is introduced, $D^i_j$, and using appropriate clauses the relation between $D^i_j$, $x_i$, $D^{i-1}_j$ and $D^{i-1}_{j-a_i}$ are described.
\[\hspace{1cm} D_{i}^{j}=\ \ \left\{
  \begin{array}{l l}
    \mbox{$\top$} & \quad \mbox{if $i$ and $j$ are both zero};\\
    \mbox{$\bot$} & \quad \mbox{$i=0$ and $j> 0$};\\
    \mbox{$(D_{j-a_i}^{i-1} \land x_{i})\lor (D_j^{i-1}\land \lnot x_{i})$} & \quad Otherwise
  \end{array}
\right. \]

  Describing $D^i_j$ variables in a top-down manner, as proposed by \cite{bailleux2006translation}, usually generate fewer number of Tseitin variables and smaller CNF than the bottom-up procedure. Translation through BDD is generalized arc-consistent but it might produce an exponential size CNF with respect to the input size.

\subsection{Non-Arc-consistent Encodings}

\subsubsection{Binary Encoding (Bin)}
  Every circuit can be translated into CNF, and so the binary adders can be described using a series of clauses. The main idea in this approach is to use binary encoding of integers and using the fact that setting $x_i$ to false is the same as setting $a_i$ to zero. Every coefficient in a PB-constraint, $a_i$, is represented as a vector of bits $\langle c^1_i\land x_i,\cdots, c^k_i\land x_i\rangle$ and each of these vectors is fed into an adder-network. The output of the adder-network is compared with the binary representation of $b$.
  
  The size of CNF generated using this encoding is polynomial with respect to the size of input but unit propagation performs poorly on the produced CNF.

  \subsubsection{Translation Through Totalizer}
  In \cite{bailleux2009new}, the authors described an encoding for PB-constraints in form $Q:\sum a_ix_i< b$ which fully supports generalized arc-consistency and produces a polynomial size CNF. In their context, setting a variable from $X$ to false never makes the constraint inconsistent, i.e., the formula $\lnot Q$ is {\em a monotone formula} \cite{MonotonicityDefinition}.

  They used gadgets, called \textit{polynomial watchdog}. A polynomial watchdog associated with the constraint $Q$ on variables $X$ is a CNF formula, $PW(Q)$, such that for every partial assignment to the input variables, $X$, that violates the constraint $Q$, unit propagation applied to $PW (Q)$ infers the value \True\ for the output variable of $PW (Q)$. 

  If constraint $Q:\sum a_ix_i<b$ is not satisfiable under a partial assignment, the sum of coefficients of variables which are set to true under the current partial assignment should be greater than or equal to $b$. The variable $x_k$ is forced to be false under current assignment iff $Q_k:\sum_{i\ne k} a_ix_i<b-a_i$, is not consistent. Global polynomial watchdog, GPW, and Local polynomial watchdogs, LPW, are used to enable UP to do these kinds of inferences. The following can be used as an encoding for PB-constraint $Q$: 

$$F=\lnot GPW(Q) \land\bigwedge LPW(Q_k)\Rightarrow (\lnot x_k).$$

Having access to an encoding for PB-constraints in the form $Q:\sum a'_il_i< b'$, one can built an encoding for constraint $Q':\sum a_il_i= b$ using the following observation:

\begin{myobservation} The set of solutions to (\ref{Q1}) is the same as the intersection of sets of solutions to (\ref{Q2}) and (\ref{Q3})
  \begin{compactenum}
    \item\label{Q1} $\sum a_il_i=b$
    \item\label{Q2} $\sum a_il_i<b+1$
    \item\label{Q3} $\sum a_i\lnot l_i<\sum a_i+ 1-b$
  \end{compactenum}
\end{myobservation}

  There are normalized PB-constraints for which totalizer based translation is not arc-consistent but our encoding is. We characterized these instances in section \ref{GAC}.

\subsubsection{Translation Through Network of Sorters (SN)}
A sorting network is a circuit with $n$ input wires and $n$ output wires consisting of a set of comparators with two input wires and two output wires. Each output of a comparator is used as an input to another comparator except those used as output wires of the sorting network. 

  In this translation, a mixed-base, $B=\langle B_1,\cdots,B_k\rangle$ is selected. And each coefficient, $a_i$, is represented using a vector of size $k$, $\langle c^1_i,\cdots,c^k_i\rangle$ such that $0\le c^j_i< B_j$ and

$$a_i=\sum_{j=1}^k c^j_i\prod_{k=1}^{j-1} B_k$$

Then each digit, $c^j_i$, is represented using $B_j$ bits (in unary encoding). $k$ sorting networks are used to implement an adder-circuit which computes the summation of $(a_i \land x_i)$ for $i=1\cdots n$. One can find more details about the translation using a network of sorters in \cite{Een2006translating}.

  The size of the CNF generated using this encoding is polynomial with respect to the size of input. This encoding is arc-consistent if all the coefficients are one. This special class of PB-constraints is called \textit{Cardinality Constraint} in SAT community. There are some well-known encodings for cardinality constraints which are arc-consistent and produce smaller CNFs \cite{Aavani2010Encoding}.

\subsection{Summary}

  Table \ref{SummaryOfApproaches} summarizes the number of auxiliary variables, clauses, and literals produced by each approach in the translation of $a_1l_1+\cdots+a_nl_n=b$. 

  BDD encoding is the only encoding which is generalized arc-consistent for this kind of PB-constraint. This encoding may produce exponential size CNF. 

 We show in section \ref{GAC} that Totalizer encoding is not arc-consistent for all constraints whose comparison operator is `$=$'. The translation using sorting networks has a reasonable size but it is arc-consistent if all the coefficients are equal to one (The authors in \cite{Een2006translating} demonstrated a \textit{necessary condition} for arc-consistency). Our encoding, equipped with $\tt M^p$ as the set of modulos and BDD translation for PBMod-constraints as translation oracle, produces a polynomial size CNF. In the next section, we show that the number of instances for which the CNF obtained by the proposed encoding is generalized arc-consistent is much more than that of sorting networks. And there are many instances for which our encoding is arc-consistent while totalizer-based encoding is not.

\begin{table}\caption{Performance of Translations ($a_{Max}=Max\{a_i\}$)}
\begin{tabular}{llll}\label{SummaryOfApproaches}
 &  \# of Auxiliary Vars.& \# of Clauses & Size of CNF\\
BDD & $O(n^2a_{Max})$ & $O(n^2a_{Max})$ & $O(nb)$ \\
Totalizer & $O(n^2\log n\log a_{Max})$ & $ O(n^3\log n\log a_{Max})$ & $O(n^3\log n\log a_{Max})$\\
Bin & $O(n\log a_{Max})$ & $O(n\log a_{Max})$ & $O(n\log a_{Max})$ \\
SN &  $O(n\log a_{Max}\log^2\log a_{Max})$ & $O(n\log a_{Max} \log^2 \log a_{Max})$ & $O(n\log a_{Max} \log^2 a_{Max})$ \\
Proposed &  $O (n(\log n+\log a_{Max})^2)$ & $ O (n(\log n+\log a_{Max})^2)$ & $ O (n(\log n+\log a_{Max})^2)$
\end{tabular}
\end{table}

  In summary, Totalizer-based encoding, Sorting Network encoding and our encoding produce polynomial size translations for PB-constraint in form $\sum a_il_i=b$ and each of them is arc-consistent for a certain subset of all possible PB-constraints. 

%

\section{Performance of Unit Propagation }\label{GAC}

  In this section, we show that there cannot be an encoding for PB-constraint in form $\sum a_il_i=b$ which always produces a polynomial size arc-consistent CNF. Also we study the arc-consistency of our encoding as well as that of Sorting Network and Totalizer encodings.

\subsection{Hardness Result}
  Here, we show that it is not very likely to have an arc-consistent encoding which always produces polynomial size CNF.

\begin{mytheorem} There does not exist a UP-detectable encoding which always produces polynomial size CNF unless P= CONP. There does not exists a UP-maintainable encoding which always produces polynomial size CNF unless P= CoNP.
\end{mytheorem}
{\noindent \bf Proof}
  Unit propagation, on a set of clauses, completes its execution either by reporting inconsistency or eliminating some variables from the input CNF. The worst-case running time of unit propagation is polynomial in size of the input CNF. 

  The subset sum problem is: given a set of integers $A=\{a_1,\cdots,a_n\}$ and an integer $b$, does the sum of a non-empty subset equal to $b$? This problem can be represented as the following PB-constraint:

  $$Q: a_1x_1+\cdots+a_nx_n=b$$

  We know that the subset sum problem is an NP-complete problem. Now, assume there exists an encoding whose resulting CNF is UP-detectable for all PB-constraints in the form $\sum a_il_i=b$. Let's call this encoding $E$. Based on definition of UP-detectability, $E$ gets a PB-constraint $Q$ and returns a valid translation $\langle v,C\rangle$ such that 

$$Q\models \bot \Leftrightarrow v\land C\sststile{UP}{} \bot.$$
 The formula $Q\models \bot$ asserts that $Q$ is not satisfiable, i.e., the original subset sum problem does not have any solution. The fact that UP can infer inconsistency on $v\land C$ in polynomial time with respect to the number of literals in $\{v\}\cup C$ implies that if $C$ has polynomial size, with respect to $Q$, deciding if the answer to a subset sum instance is `No' is easy. That is, either there are PB-constraints whose corresponding CNFs are not polynomial size or CoNP=P.

  Now, consider the following problem: Given a normalized PB-constraint $Q:\sum a_il_i=b$, does it have exactly one solution?

  The Unique SAT problem, USAT, can be reduced to this problem. The reduction is similar to the reduction explained in \cite{sipser1996introduction} to prove the NP-hardness of subset sum problem (we did not include it in this paper for sake of space). It is already known that USAT belongs to complexity class $D^P$ and it is CoNP-hard \cite{papadimitriou1982complexity}. 

  Let $Q:\sum a_ix_i=b$ be the output of the reduction on the USAT instance $C$. $C$ has exactly one solution iff $Q$ has exactly one solution. But if $Q$ has exactly one solution, \cA, we have $Q\models x_i$ iff \cA$\models x_i$ and $Q\models \lnot x_i$ iff \cA$\not\models x_i$. Let $\langle v,C\rangle$ be a UP-inferable translation for $Q$, then we should have
\begin{equation}
  \forall i: {\cal A}\models x_i: C\land v\sststile{UP}{} x_i\\
  \forall i: {\cal A}\not\models x_i: C\land v\sststile{UP}{} \lnot x_i
\end{equation}
  So, UP can infer all input variables values, $x_1,\cdots,x_n$, when it is executed on $C\land v$ iff the given subset sum instance has exactly one solution.

  Throughout the rest of this section, we assume we are given a PB-constraint, $Q:\sum a_il_i=b$ and a valid translation for it, $\langle v,C\rangle$. Also, let $Q_1,\cdots,Q_m$ be the PBMod-constraints generated during the translation process and $\langle v_i,C_i\rangle$ be a valid translation for $Q_i$. Also, assume $Ans=\{A_1,\cdots,A_r\}$ is the set of all possible solutions to $Q$.

\subsection{Arc-consistency for Proposed Encoding}

\noindent  There are three situations in which UP is able to infer the input variables values and so one can expect SAT solvers to perform well in those situations:
  \begin{compactenum}
  \item{Unit Propagation Detects Inconsistency}:
   One can infer there is no assignment satisfying $Q$ by knowing $Ans=\emptyset$. We call the unsatisfiable constraints whose translations are UP-detectable to be \textit{good constraints}.


  UP gets $\{v\}\cup C$ as its input, it detects $v$ should be true and next, it finds out $v_i$ is true, for all $1\le i\le m$. Based on Proposition \ref{ArcConsistencyDPPBMod}, if at least one of the $m$ PBMod-constraints is unsatisfiable, UP detects inconsistency. 

  \item{Unit Propagation Solves Constraint}:
   One can infer the solution for $Q$ if there is just a single satisfying solution to $Q$, i.e., $Ans=\{A_1\}$. For this kind of constraints, UP might be able to infer the correct values for all input variables ($X$). We call the constraints which have exactly one solution and UP is able to solve them completely the \textit{nice constraints}. Note that after a consistent solution to the input variables has been found, the values of all auxiliary variables generated during the translation are either forced or `don't care'.

  UP gets $\{v\}\cup C$ as its input, it detects $v$ should be true and next, it finds out $v_i$ is true, for all $1\le i\le m$. Based on Proposition \ref{ArcConsistencyDPPBMod}, if at least one of the $m$ PBMod-constraints has exactly one solution, UP is able to infer all the input variables value.

  \item{Unit Propagation Infers the Value for an Input Variable}:
   One can infer the value of input variable $x_k$ is \True/\False\ if $x_k$ takes the same value in all the solutions to $Q$. For this kind of constraints, UP might be able to infer the value of $x_k$. Note that the nice constraints are a subset of these constraints.

  Similar to case of nice constraints, UP detects $v$ should be true and next, it finds out $v_i$ is true, for all $1\le i\le m$. Based on Proposition \ref{ArcConsistencyDPPBMod}, UP infers the correct value for $x_k$ if $x_k$ has the same value in all of solutions to at least one of the $m$ PBMod-constraints. 

  \end{compactenum}

These three cases are illustrated in the following example.
\begin{myexample} In this example, we use the same PB-constraint as we used in Example \ref{Example1}.
  \begin{compactenum} 
      \item If $A$, the current partial assignment, is $A=\{\lnot x_2,\lnot x_4\}$ and $P=5$. There is no total assignment satisfying $1x_1+3x_3=2$.
      \item If $A$, the current partial assignment, is $A=\{x_2,\lnot x_3,x_5\}$ and $P=3$, there is exactly one total assignment ($\{\lnot x_1,x_2,\lnot x_3,\lnot x_4,x_5\}$) which extends $A$ and satisfies the PBMod-constraint. 
%
%
      \item If $A$, the current partial assignment, is $A=\{\lnot x_3,\lnot x_5\}$ and $P=2$, there are four total assignments extending $A$ and satisfying the PBMod-constraint. In all of them, $x_1$ is in the solution. 
%
  \end{compactenum}
\end{myexample}

\noindent In the rest of this section, we estimate the number of good and nice constraints, i.e., we give a lower bound for the number of constraints whose translation can be solved just by using unit propagation.

  Let us assume the constraints are selected, uniformly at random, from $\{\sum a_1l_1+\cdots+a_nl_n=b:1\le a_i\le A=2^{R(n)} \textrm{ and } 1\le b\le n*A\}$ where $R(n)$ is a polynomial in $n$ and $R(n)>n$. To simplify the analysis, we use the same prime modulos ${\tt M^P}=\{P_1=2,\cdots, P_m=\theta (R(n))>2n\}$ for all possible constraints. 
 
  Consider the following PBMod-constraints:

\begin{eqnarray}
\label{NoSolution}  1x_1+\cdots+ 1x_{n-1}+ 1x_{n}= n+1 (\textrm { mod }P_m), \\
\label{OneSolution} 1x_1+\cdots+ 1x_{n-1}+ 2x_{n}= n+1 (\textrm { mod }P_m), \\
\label{OneFixedValue} 1x_1+\cdots+ 1x_{n-1}+nx_{n}= 2n-2 (\textrm { mod }P_m). 
\end{eqnarray}

  One can verify that (\ref{NoSolution}) does not have any solution, (\ref{OneSolution}) has exactly one solution and $x_n$ is \True\ in all solutions for (\ref{OneFixedValue}). {\em Chinese Remainder Theorem}, \cite{ChineseRemainderTheoremRef}, implies that there are $(A/P_m)^{n+1}=2^{(n+1)Q(n)}/Q(n)^{n+1}$ different PB-constraints in the form $\sum a_1l_i=b$ such that their corresponding PBMod-constraints, where the modulo is $P_m$, are the same as (\ref{NoSolution}). The same thing is true for(\ref{OneSolution}) and (\ref{OneFixedValue}). 

\subsection{UP for Sorting Network}
  Here, we show that there are more instances for which our encoding maintains arc-consistency than Sorting Network. 

  It is stated in \cite{Een2006translating}:
 ``Unfortunately, arc-consistency is broken by the duplication of inputs, both to the same sorter and between sorters."

As we described in Section \ref{PreviousWork}, in Sorting Network encoding, one fixes a multi-base $B=\langle B_1,\cdots,B_m\rangle$. To avoid duplication between sorters, each coefficients, $a_i$, should have a single non-zero digit in their multi-base $B$-representation. To avoid duplication in the same sorter, the non-zero digit should be exactly 1. So, each coefficient can take $m$ different values, based on the position of its non-zero digit. There are $n$ coefficients, so there are at most $nAm^n$ different instances which are arc-consistent, where $nA$ is the maximum number of possible right-hand side of the equation. Having $B_i\ge 2$ implies that $m\le \log A$. 

  \subsection{UP for Totalizer-based Encoding}
    In \cite{bailleux2009new}, it is claimed that, totalizer-based encoding is a polynomial size CNF encoding such that generalized arc-consistency is maintained through unit propagation for all PB-constraints in the following form:
  $$\sum a_il_i \{=,>,\ge,<,\le\} b.$$
  Although totalizer-based encoding is generalized arc-consistent for the PB-constraints in the forms $\sum a_il_i\{>,\ge,<,\le\}$, it does not produce an arc-consistent translation for some PB-constraints in the form $\sum a_il_i=b$. 

  In their approach, the PB-constraint $Q:\sum a_il_i=b$ should be converted to the following two constraints:
\begin{equation}
  \sum a_il_i<b+1,\\
  \sum a_i\lnot l_i<\sum a_i+1-b.
\end{equation}

  Consider the following PB-constraint:
  $$Q_1:3x_1+3x_2+4x_3=7,$$
  As $Q_1(3):0x_1+0x_2+1x_3=1$, UP, and also our approach, can infer that $x_3$ should be true. Now consider the following two constraints: 
  $$Q_2:3x_1+3x_2+4x_3<8,$$
  $$Q_3:3\lnot x_1+3\lnot x_2+4\lnot x_3<10-6=4.$$

  Let $\langle v_2,C_2\rangle$ and $\langle v_3,C_3\rangle$ be valid translations obtained from totalizer-based encoding for $Q_2$ and $Q_3$, respectively. UP does not infer anything from $v_2\land C_2$ because nothing can be inferred about any of $x_i$s by knowing $Q_2$ should be true. We have the same situation for $Q_3$.

  In fact, the translation produced by totalizer-based encoding is not generalized arc-consistent for almost all PB-constraints which have a PBMod-constraint in form (11) or (10).

We summarize the discussion above in the following observations:
\begin{myobservation}  There are at most $(\log A)^n$ instances where the CNF produced by Sorting Network encoding maintains arc-consistency, while this number for our encoding is at least $(A/\log (A)))^n$. So, if $A=2^{R (n)}$, almost always we have $2^{R(n)}/R(n)\gg R (n)$.
\end{myobservation}

\begin{myobservation}  There is a family of PB-constraints whose translation through totalizer-based encoding is not arc-consistent but the translation obtained by our encoding is arc-consistent. 
\end{myobservation}

\section{Conclusion and Future Work}\label{Conclusion}

  We presented a method for translating Pseudo-Boolean constraints into CNF. The size of produces CNF is polynomial with respect to the input size. We also showed that for exponentially many instances, the produced CNF is arc-consistent. This number is much bigger than that of the existing encodings.
 
  The upper bounds on the size of CNF are not tight. One needs to analyze the performance of the proposed method more carefully and find a tighter bounds on the CNF size. We still need to implement the proposed encoding and compare it with the other encodings on some real-life problems. 




\begin{small}
\bibliographystyle{plain}
\bibliography{MyPhDBib}
\end{small}

\appendix
\section{Proposition 7 (Proof)}\label{Proposition7Proof}

Here, we prove Proposition (\ref{BoundOnM}) presented in section \ref{ProposedMethod}. 

Let ${\tt M^P}$ be the set of $m$ first prime numbers, ${\tt M^P}= \{2, 3, \cdots, P_m\}$ and $S$ be an integer. 

  Prime number theorem, \cite{PrimeNumberTheoryRef}, states that the number of prime number less than or equal to an integer $x$, $\pi(x)$, satisfies the following:

\begin{equation}
\label{PrimeNumberTheorem}\lim_{x\mapsto \infty} \frac{\pi(x)}{x/ln(x)}=1.
\end{equation}

Using (\ref{PrimeNumberTheorem}), we can bound the value of $\Pi_{p\in {\tt M^P}} p$, by:

\begin{eqnarray}
\label{UBLBProduct1}  \left(\frac{P_m}{e}\right)^{\pi(P_m)-\pi(P_m/e)}\le \prod_{p\in {\tt M^p}} p \le \left(P_m\right)^{\pi(P_m)}
\end{eqnarray}

 By setting $\pi(x)=x\slash\ln x$, we can rewrite (\ref{UBLBProduct1}) as:

\begin{eqnarray}
\label{UBLBProduct2} \left(\frac{P_m}{e}\right)^{P_m/\ln (P_m)-P_m/\left(e*\ln (P_m/e)\right)}\le \prod_{p\in {\tt M^p}} p \le \left(P_m\right)^{P_m/\ln (P_m)}\le e^{P_m}
\end{eqnarray}

A lower bound for $\Pi_{p\in {\tt M^P}}$ can be obtained as follows:

\begin{eqnarray}
\nonumber  \left(\frac{P_m}{e}\right)^{P_m/\ln (P_m)-P_m/\left(e*\ln (P_m/e)\right)} \\
 \label{LBProduct} = \left(e\right)^{\frac{P_m* \left(\ln P_m- 1\right)}{\ln P_m}- \frac{P_m}{e}}\ge  \left(e\right)^{\frac{P_m}{2}- \frac{P_m}{e}}
\end{eqnarray}

From (\ref{UBLBProduct2}) and (\ref{LBProduct}):
\begin{eqnarray}
\label{UBLBProduct} \left(e\right)^{\frac{P_m}{2}- \frac{P_m}{e}} \le \prod_{p\in {\tt M^P}} p \le e^{P_m}\\
\label{BoundProduct}\prod_{p\in {\tt M^P}} p\in e^{\theta(P_m)}
\end{eqnarray}

The last equation, (\ref{BoundProduct}), states that the maximum value in {$\tt M^P$} whose product is larger than a given $S$ is $\theta(\ln S)$.

Now, by applying the prime number theorem once more, we get that:
$$m=|{\tt M^P}|\approx \frac{P_m}{\ln P_m}= \theta(\frac{\ln S}{\ln \ln S})$$

%

\end{document}